\documentclass[sigconf]{acmart}
\renewcommand\footnotetextcopyrightpermission[1]{}
\acmConference[]{}{}{}

\usepackage{graphicx}
\usepackage{enumitem}
\usepackage{tabularx}
\usepackage{booktabs}
\usepackage{hyperref}
\usepackage{subcaption}
\usepackage[most]{tcolorbox}
\usepackage{pifont}
\usepackage{fontawesome}
\usepackage{xurl}   
\usepackage{xcolor}
\usepackage{placeins}
\usepackage{float}

\definecolor{findingbg}{HTML}{F5F7FA}
\definecolor{findingaccent}{HTML}{0078B8}
\definecolor{findingtext}{HTML}{FFFFFF}

\definecolor{gapoppbg}{HTML}{FFF9E6}
\definecolor{gapoppaccent}{HTML}{B8860B}

\newtcolorbox{gapoppbox}[1][]{
  enhanced,
  colback=gapoppbg,
  colframe=gapoppaccent,
  boxrule=0.5pt,
  sharp corners,
  frame hidden,
  boxrule=0pt,
  borderline west={2pt}{0pt}{gapoppaccent},
  left=4pt,
  right=4pt,
  top=2pt,
  bottom=2pt,
  #1
}

\newtcolorbox{findingbox}[2][]{
  enhanced,
  colback=findingbg,
  colframe=findingaccent,
  boxrule=0pt,
  borderline west={2pt}{0pt}{findingaccent},
  sharp corners,
  left=4pt,
  right=4pt,
  top=2pt,
  bottom=2pt,
  fonttitle=\bfseries\color{findingtext},
  title={\(\triangleright\) #2},
  #1
}

\begin{document}

\title{Characterizing Datasets for LLM-based Requirements Engineering: A Systematic Mapping Study}

\author{Quim Motger}
\email{joaquim.motger@upc.edu}
\affiliation{%
  \institution{Universitat Politècnica de Catalunya}
  \city{Barcelona}
  \country{Spain}
}

\author{Carlota Catot}
\email{carlota.catot@upc.edu}
\affiliation{%
  \institution{Universitat Politècnica de Catalunya}
  \city{Barcelona}
  \country{Spain}
}

\author{Xavier Franch}
\email{xavier.franch@upc.edu}
\affiliation{%
  \institution{Universitat Politècnica de Catalunya}
  \city{Barcelona}
  \country{Spain}
}

\renewcommand{\shortauthors}{Motger et al.}



\acmSubmissionID{Preprint}

\begin{abstract}
Large Language Models (LLMs) depend on high-quality, domain-specific natural language datasets. This dependency is particularly pronounced in Requirements Engineering (RE), where core activities rely on textual artifacts such as requirements, specifications, and stakeholder feedback. Despite the increasing use of LLMs in RE, data scarcity remains a widely reported limitation.
While several datasets support LLM-based RE research, they are scattered across studies and lack systematic characterization, hindering reuse, comparability and assessment. This paper addresses this gap by examining which public datasets are used in LLM-based RE, how they can be consistently characterized, and which RE tasks and dataset properties remain under-represented.
We report on a systematic mapping study of 45 primary studies referencing 62 publicly available datasets. Each dataset is characterized using a structured scheme covering multiple dimensions, including relevant descriptors such as artifact type, granularity, RE activity, supported task, application domain, and language, among others. 
The results reveal notable imbalances, including an incomplete adoption of open-science practices, limited dataset support for elicitation activities, and a lack of language and socio-technical diversity. 
The resulting catalogue and characterisation scheme support informed dataset selection, comparison, and reuse, contributing to stronger empirical foundations for LLM-based RE research and evaluation.
\end{abstract}

\maketitle              

\section{Introduction}
\label{sec:introduction}

The need for data in the era of Large Language Models (LLMs) has become imperative. Domain-specific datasets are essential to increase LLM task performance at multiple stages, including pre-training, fine-tuning, in-context learning, model inference, and evaluation~\cite{Hou2024}. Despite this, data scarcity is reported as a common limitation in recent requirements engineering (RE) surveys and mapping studies~\cite{Zhao2021,Zadenoori2025}. While popular RE datasets such as PURE~\cite{ferrari2018pure} or PROMISE~\cite{clelandhuang2007nfr} are repeatedly referenced, used, and extended, RE researchers and practitioners face this limitation across multiple dimensions, including different RE activities (e.g., elicitation, specification, validation), tasks (e.g., classification, traceability, term extraction), domains (e.g., education, healthcare, mobility), artifact types (e.g., requirements, user stories, user reviews), and other dataset attributes. Moreover, many datasets remain confined to the context of individual research groups or publications, with limited reuse or cross-project integration. Hence, navigating across the spectrum of available datasets becomes challenging, especially when multiple dimensions come into play. Consequently, identification and characterization of the current state of research on datasets used for LLM-based RE tasks (LLM4RE), as well as research gaps and opportunities to contribute, remains obscure.

The RE community has explored dataset availability through literature reviews~\cite{Hou2024,Zadenoori2025}, but these typically focus on narrowed dataset identification without deeper characterization or comparison. Open science initiatives provide promising directions: knowledge bases for empirical RE~\cite{Karras2023} leverage open source initiatives such as the Open Research Knowledge Graph (ORKG)~\cite{Stocker2023FAIRRKG} to improve discoverability and reusability. Yet, they depend on proactive contributions and remain more research-oriented than dataset-oriented. Dedicated dataset efforts, such as the ORKG datasets~\cite{ORKGDatasetContentType} and the RE Open Data Initiative~\cite{ZenodoREOpenDataInitiative}, partially address these gaps by providing public repositories for RE specific datasets. However, the number of available datasets remains severely limited, contributions must be manually submitted by researchers, and systematic characterization and cross-dataset analysis are still lacking.

To address these challenges, we conducted a systematic mapping study of publicly available datasets used in LLM4RE research across all RE stages. Our contributions are:

\begin{itemize}
    \item A public catalogue of 62 RE datasets\footnote{Catalogue available at \href{https://nlp4se.github.io/LLM4RE-Datasets/}{https://nlp4se.github.io/LLM4RE-Datasets/}} stemming from 45 studies. 
    \item A reproducible methodology and protocol for systematic mapping of RE datasets, enabling iterative updates to integrate new datasets and continuously expand the catalogue.
    \item Characterization and empirical analysis of the datasets, considering both quantitative and qualitative aspects based on generic and RE-specific descriptors.
    \item Elicitation of research gaps and opportunities in dataset availability, coverage, quality, and reuse, providing directions for future LLM4RE research and dataset development.
\end{itemize}
\section{Background \& Related Work}
\label{sec:rel-work}

There are several literature reviews and mapping studies on Natural Language Processing (NLP) for RE (NLP4RE) published in recent years. Zhao et al. conducted one of the largest literature reviews in the field~\cite{Zhao2021}, prior to the disruption brought by LLMs and generative artificial intelligence. In this study, the authors already pointed out the scarcity of data, urging that \emph{NLP4RE researchers share RE-specific datasets, benchmark data, and performance metrics}~\cite{Zhao2021}. Necula et al. complemented this vision with a more up-to-date perspective on LLM-based tasks~\cite{Necula2024}. With the rise and adoption of LLMs in research, initial studies started to categorize LLM4RE research leveraging ChatGPT~\cite{Ronanki2023,Marques2024} and other LLMs~\cite{Fan2023,Khan2024,Vogelsang2025}. While all of these studies highlight the potential of using domain-specific datasets in LLM-based tasks, none of them report on the specific datasets used in primary studies.

As LLM-based research consolidates, several studies have started to address dataset availability as a study dimension in this domain. Hou et al. conducted one of the largest, most recent literature reviews on LLM-based research within the context of software engineering~\cite{Hou2024}. While they reported up to 374 datasets, the vast majority pertain to code-based datasets (e.g., code, programming tasks/problems, programming prompts), and they do not provide a characterization of tasks and activities within the context of RE. Beyond generic reviews, dedicated studies address the dataset dimension for particular software- and requirements- engineering tasks. H\"{a}m\"{a}l\"{a}inen et al. conducted a systematic literature review on multi-label research and datasets in software engineering~\cite{Hamalainen2025} , with a particular focus in RE. Specifically, they identified 7 studies on text classification tasks of multiple document types, such as issues~\cite{AlDhafer2022}, requirements~\cite{ALHOSHAN2023107202} and software change requests~\cite{Ahsan2010}. Although these works analyse dataset descriptors such as data types and reuse, their analysis is conducted at the software engineering level rather than at the RE level. Moreover, their scope extends beyond LLM-based approaches to include traditional machine learning methods. While such breadth may reveal additional datasets potentially relevant to LLM4RE, it also broadens the analysis beyond our objective. In contrast, our study deliberately focuses on datasets whose value has already been demonstrated within LLM4RE research, enabling a more precise characterization of datasets that have been empirically validated in this specific context.

Tailored to our scope of analysis, Cheng et al. conducted a systematic literature review on generative AI for requirements engineering~\cite{Cheng2025GenAI4RE}. They address evaluation methods, metrics, tools and datasets in LLM4RE research throughout a dedicated research question. However, with respect to datasets, their analysis is limited towards availability, claiming that less than half of the surveyed studies (48.7\%) provide available datasets. Despite this finding, which further motivates our study, no additional analysis is provided.
More recently, Zadenoori et al. conducted a systematic literature review of LLM4RE research~\cite{Zadenoori2025}. Their study includes a proposal for a domain-specific conceptual scheme for data extraction, including specific vocabulary for attributes such as \textit{RE task} and \textit{RE stage}, which we reuse and adapt in our study for consistency and traceability. While their characterization is tailored to the RE domain, only 10 publicly available datasets are reported. Conversely, our research characterizes 62 publicly available datasets, including 9 out of the 10 datasets identified by Zadenoori et al.~\cite{Zadenoori2025}, excluding LeetCode, which is classified as \textit{Software Engineering Tasks} and thus out of scope.

Consequently, and to the best of our knowledge, no prior study provides a comprehensive, RE-specific characterization of publicly available datasets validated in LLM4RE research. Existing secondary studies mainly identify datasets and categorize them by RE activity, source, or artifact type, but overlook key aspects such as licensing, accessibility, size, labels, granularity, domain, and lifecycle coverage. Moreover, the value of consolidating datasets into a unified catalogue remains underexplored. This study addresses these gaps through a structured, dataset-centric characterization and a centralized catalogue. 
\section{Study Design}
\label{sec:design}

\subsection{Research Questions}

The goal of this study is to \textbf{provide a structured characterization of public datasets for LLM-based tasks in RE}. The results aim to support researchers and practitioners in identifying, comparing, and selecting suitable datasets. To this end, we formulate the following research questions (RQs):

\begin{enumerate}[label=\textbf{RQ\arabic*}]
\item How do LLM4RE datasets vary in terms of provenance, accessibility and reuse?
\item How are LLM4RE datasets distributed across quantitative and general-purpose dataset descriptors?
\item How are LLM4RE datasets applied within the requirements engineering lifecycle?
\item What research gaps and opportunities for LLM4RE research arise from the current distribution of dataset descriptors?
\end{enumerate}

RQ1 examines the maturity and sustainability of LLM4RE dataset practices by analysing publication timelines, hosting platforms, licensing, and dataset extensions, to reveal trends in availability, openness, and reuse practices.
RQ2 examines quantitative and general-purpose descriptors, such as domain, size, language, and artifact granularity, enabling the identification of patterns in dataset coverage, diversity, and scalability. 
RQ3 considers RE-specific descriptors, including document type, RE lifecycle activity, NLP task, and complementary semantic information such as annotated labels, providing insights into how datasets support different RE tasks and lifecycle stages. 
Finally, RQ4 integrates the findings from RQ1–RQ3 to identify overarching limitations and gaps, highlighting directions for future dataset creation and research in LLM4RE.

\subsection{Method}
\label{sec:method}

\begin{table*}[h]
\centering
\caption{Data extraction descriptors for each dataset entry, with value types (PB = Pattern-based, OV = Open vocabulary, CV = Closed vocabulary) and example values from the PROMISE\_exp dataset.}
\begin{tabularx}{\textwidth}{p{1.6cm} p{4.5cm} p{4.5cm} X}
\toprule
\textbf{Field} & \textbf{Description} & \textbf{Type} & \textbf{Example (PROMISE\_exp)} \\
\midrule
Code & Unique dataset identifier used in this review. & \textit{(PB)} Pattern-based ID (D\textit{XYZ}) & D044 \\
Name & Short dataset name; if missing, we use first author’s last name and year. & \textit{(OV)} Short string & PROMISE\_exp \\
Description & Summary of the dataset content and purpose. & \textit{(OV)} Free text & ``Extension from PROMISE dataset containing requirements classified into functional and non-functional sub-types" \\
Reference & Full citation to the dataset source or accompanying publication. & \textit{(OV)} Citation string (APA) & Lima et al. Software engineering repositories: Expanding the PROMISE database. In Proceedings of the 33rd Brazilian Symposium on Software Engineering, Salvador, Brazil, 23–27 September 2019; pp. 427–436. \\
Year & Release year of the dataset. & \textit{(PB)} 4-digit integer & 2019 \\
URL & Persistent link to the dataset. & \textit{(PB)} URL & \url{https://github.com/AleksandarMitrevski/se-requirements-classification} \\
License & Dataset license information. & \textit{(CV)} SPDX code or textual label & None \\
Artifact & Type of textual artifact represented. & \textit{(CV)} bug, QA, requirements, user reviews, legal text, post, questions, user stories, other & requirements \\
Granularity & Unit of annotation or analysis. & \textit{(CV)} document, sentence, pair, token, other & document \\
Activity & Targeted RE stage or primary usage. & \textit{(CV)} elicitation, analysis, specification, management, verification \& validation, other & analysis \\
Task & Primary computational task supported. & \textit{(CV)} classification, extraction, traceability, modelling, Q\&A, other & classification \\
Domains & Application areas of the dataset & \textit{(PB)} List of short strings & security, real state, healthcare, finance, ...
 \\
Size & Number of textual artifacts. & \textit{(PB)} Positive integer & 969 \\
Languages & Languages used in the dataset. & \textit{(PB)} ISO~639-1 codes & en \\
Labels & Classes or tags used in the dataset. & \textit{(OV)} Text labels & Functional, Availability, Legal, Look and feel, Maintainability, Operational, Performance, Scalability, Security, Usability, Fault tolerance, Portability \\
Extends & Dataset(s) this entry extends or derives from. & \textit{(PB)} \textit{Code} values & D043 \\
Publications & Publications using this dataset covered in this study. & \textit{(OV)} Publication IDs & P019, P028, P052, P120 \\
\bottomrule
\end{tabularx}
\label{tab:extraction}
\end{table*}

We follow the methodology of systematic mapping studies proposed by Petersen et al.~\cite{Petersen2008,PETERSEN20151}, which defines the planning, conducting, and documenting phases, and which is widely adopted in software engineering and RE research. Complementarily, and inspired by similar studies~\cite{Annunziata2024,Williams2022,ARDIC2025112447}, we followed the ACM/SIGSOFT Empirical Standards for conducting and reporting systematic reviews~\cite{Ralph2021}.

\textbf{Scope and definitions.} We define the \emph{target phenomenon} as the use of public datasets for LLM-based tasks operating on RE textual artifacts (i.e., written in natural language) across RE lifecycle activities. 
A \emph{dataset} is any publicly accessible corpus providing a stable reference or URL and sufficient metadata to enable reuse~\cite{Bosu2013}. 
\emph{LLM-based tasks} include pre-training, fine-tuning, in-context learning, and downstream inference, benchmarking, and evaluation for RE-related activities~\cite{Hou2024}. 
\emph{RE textual artifacts} encompass requirement-related documents, such as textual requirements, user stories, issue reports, app reviews, specifications, models, test cases, and trace links~\cite{Franch2023}.
Finally, \emph{RE activities} cover elicitation, analysis, specification, management, verification and validation stages~\cite{Zadenoori2025}. 
These definitions not only delimit the scope of the study for constructing the search string but also inform the design of data extraction using domain-specific vocabularies for systematic characterization.

\textbf{Databases.} Following recent systematic mapping studies in the software engineering community~\cite{Annunziata2024}, we primarily rely on Scopus as the main source of white literature due to its broad and comprehensive indexing of peer-reviewed research in software engineering~\cite{Martin-Martin2021}. To further increase coverage and mitigate potential selection bias, we complement this search with major publisher databases commonly used in related studies~\cite{Stevanetic2015,Aljedaani2021}, namely IEEE Xplore, ACM Digital Library and Springer Nature Link. We intentionally exclude grey literature and dataset repositories from the scope of this study to focus on peer-reviewed studies, ensuring consistent quality and replicability of our analysis.

\begin{figure*}[htbp!]
  \centering
  \includegraphics[width=\textwidth]{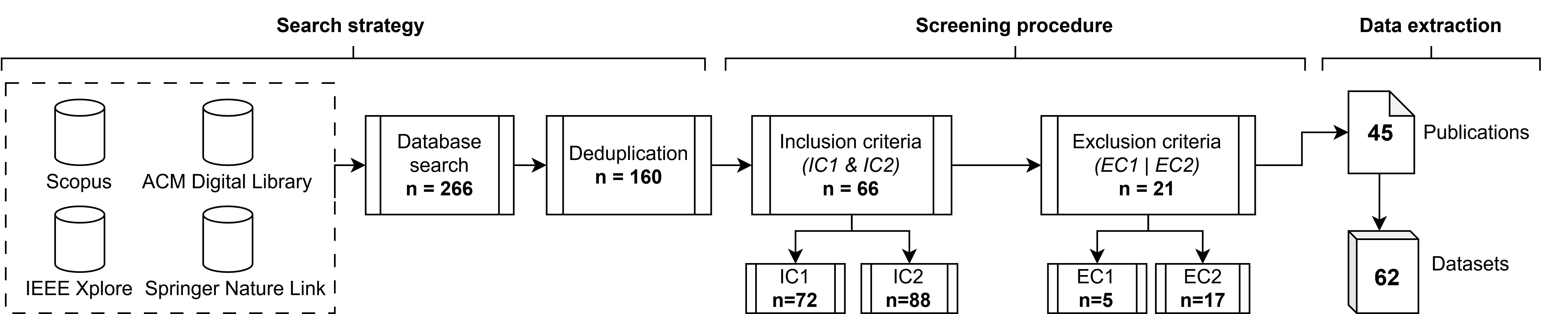}
  \caption{Study search results.}
  \label{fig:study-search}
\end{figure*}

\textbf{Search strategy.} We constructed the search string by combining three concept blocks: \texttt{LLMs}, \texttt{RE activities}, and \texttt{dataset}-related terms. The construction is informed by a PICOC perspective, where Population (LLMs), Context (RE activities), and Outcome (datasets) are explicitly captured, while Intervention and Comparison are not applicable given the exploratory nature of this mapping study. For the LLM and RE activity concepts, we based our approach on the conceptual model proposed by Zadenoori et al.~\cite{Zadenoori2025}, expanding it with synonyms and variants identified in related work (see Section~\ref{sec:rel-work}). The complete search string\footnote{The syntax is simplified for manuscript presentation and adapted to each database.} -- applied to title, abstract, and keywords -- is shown below:

\begin{center}
\begin{minipage}{\linewidth}
\small\itshape
\faDatabase\quad Search String: (``large language model" OR ``llm") AND
(``requirements" AND
(``engineering" OR ``elicitation" OR ``specification" OR ``analysis" OR
``validation" OR ``verification" OR ``prioritization" OR ``prioritisation" OR
``gathering" OR ``collection" OR ``modelling" OR ``modeling" OR
``design" OR ``traceability")) AND
(data OR corpus OR benchmark OR ``gold*standard" OR ``ground*truth")
\end{minipage}
\end{center}

The study search was conducted on September 15th, 2025, with no time-frame limitation on selected studies. All duplicates were excluded before subsequent steps.

\textbf{Inclusion/exclusion criteria.} 
We \emph{included} a study if the paper:
\begin{enumerate}[label=(IC\arabic*), leftmargin=*]  
  \item is a primary study using LLMs for RE activities,
  \item uses, releases, or refers to datasets related to RE activities.
\end{enumerate}
On the other hand, we \emph{excluded} a study if the paper:
\begin{enumerate}[label=(EC\arabic*), leftmargin=*]  
  \item is not available for review,
  \item uses or refers to datasets that are not publicly available.
\end{enumerate}

\textbf{Screening procedure.}
\label{sec:screening}
After study search, merge and deduplication, two authors of this study independently screened the original set of studies to assess inclusion and exclusion criteria. Each author reviewed a different subset of studies. To ensure consistency, both authors jointly inspected a random sample comprising 10\% of the studies, calculating inter-rater agreement. The resulting Cohen's kappa was 0.87, indicating almost perfect agreement. Disagreements were discussed and resolved before full screening.

\textbf{Data extraction schema.}
\label{sec:extraction}
Table~\ref{tab:extraction} defines the fields used for dataset characterization. 
The vocabulary was elicited from three complementary sources. 
First, we adopted generic fields (\emph{code}, \emph{name}, \emph{description}, \emph{reference}, \emph{year}) commonly used in systematic mappings~\cite{Petersen2008,Martin-Martin2021}, as well as dataset-specific fields (\emph{license}, \emph{URL}) to support provenance and accessibility analysis (RQ1). 
Additionally, the \emph{extends} field was added to capture reuse relationships between datasets (i.e., when one dataset extends another). 
Second, we incorporated data-oriented descriptors (\emph{domain}, \emph{granularity}, \emph{size}, \emph{language}) inspired by standard dataset repositories such as HuggingFace, Kaggle, and the ORKG, to support characterization of datasets (RQ2). 
Finally, NLP4RE-specific dimensions (\emph{RE activity}, \emph{task}, \emph{document type}, \emph{labels}) were adapted from recent taxonomies~\cite{Zhao2021,Zadenoori2025} to support analysis of how these datasets are applied within the requirements engineering context (RQ3). We pilot-coded 10\% of the entries and refined the codebook before full extraction. 


\textbf{Catalogue distribution.}
As a contribution of this work, and to increase the visibility, accessibility, and reuse of the study outcomes, the resulting dataset catalogue has been made available through the following dissemination channels:
\begin{itemize}
    \item \textbf{LLM4RE catalogue web application.} The curated collection of datasets is published as an interactive web-based application at \url{https://nlp4se.github.io/LLM4RE-Datasets/}. The application supports searching and filtering datasets based on the extracted descriptors, and provides a complementary dashboard offering aggregated analytical insights derived from descriptor frequencies.
    \item \textbf{ORKG comparison.} The dataset characterization is also available as an ORKG comparison~\cite{ORKGCatalogue}, which provides structured, machine-actionable overviews that enable systematic comparison of dataset descriptors through the ORKG web interface\footnote{\url{https://orkg.org/comparisons}}. This representation facilitates traceability between datasets and primary studies, supports exploratory analysis, and promotes reuse in future LLM4RE research.
\end{itemize}


\section{Provenance, Accessibility \& Reuse (RQ1)}
\label{sec:rq1}

Figure~\ref{fig:study-search} summarizes the study selection process and search results. 
After database search and deduplication, 160 distinct studies proceeded to the inclusion and exclusion criteria screening. Of these, 66 studies (41.3\%) satisfied both inclusion criteria. Among them, 21 studies (31.8\%) met at least one exclusion criterion. For EC1, studies were not available even after explicit request to the authors. For EC2, datasets were unavailable due to either motivated reasons (e.g., intellectual property or industrial confidentiality) or unmotivated reasons (e.g., missing or broken links or undocumented data sharing).
As a result, 45 primary studies were retained. Thirteen of these studies (28.9\%) used more than one dataset. Overall, the retained studies cover 62 distinct publicly available datasets.

The full list of datasets covered in this study is provided in the replication package, together with the complete per-dataset characterisation, the web-based catalogue, and the ORKG comparison. Due to space constraints, the paper reports aggregated results and  representative examples across RQ2--RQ4 to illustrate the main trends.

\begin{figure}[h]
  \centering
  \includegraphics[width=\columnwidth]{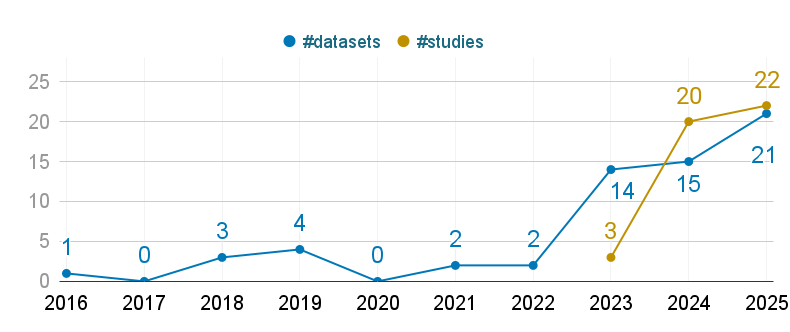}
  \caption{Publication year of LLM4RE datasets and studies.}
  \label{fig:year-distribution}
\end{figure}

Figure~\ref{fig:year-distribution} illustrates the temporal distribution of publicly available LLM4RE datasets and the publications covered in this study in which these datasets are used. While a pronounced increase is observed from 2023 onward, the figure also shows that up to 12 datasets (19.4\%) were published before the recent disruption of LLM research and have since been reused in LLM4RE studies. This indicates continuity in dataset usage, with several pre-LLM datasets being reused in later LLM-based work.

\begin{figure}[h]
  \centering
  \includegraphics[width=\columnwidth]{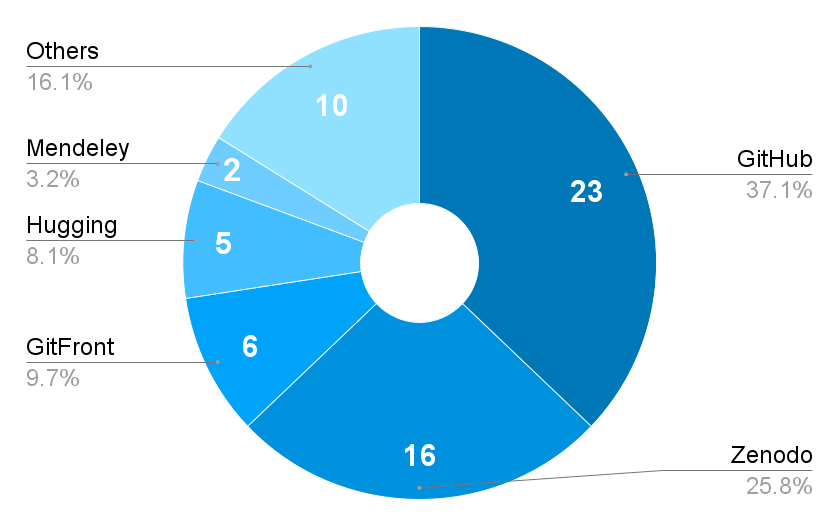}
  \caption{Distribution of dataset hosting platforms.}
  \label{fig:source}
\end{figure}

With respect to accessibility, Figure~\ref{fig:source} reports the distribution of dataset hosting platforms\footnote{Entities with only one occurrence are grouped into the \emph{Others} category.}. Together, Git-based hostings (46.8\%) are the most frequently used platforms, while research data repositories such as Zenodo, Mendeley, and FigShare are also widely adopted. These repositories typically require datasets to be assigned a persistent identifier (e.g., a DOI), fostering openness, citability, and long-term availability. Other platforms include supplementary material hosted by publishers (e.g., Springer, IEEE Xplore, Elsevier) or project-specific websites.

A cross-analysis of publication year and hosting platform reveals a gradual consolidation of dataset dissemination practices over time. Several datasets published before 2020 are hosted on project-specific websites or legacy repositories (50.0\%), whereas datasets released from 2022 onward increasingly rely on community-oriented platforms such as GitHub, Zenodo, and Hugging Face. In particular, Zenodo and Hugging Face appear almost exclusively in recent years, reflecting a growing alignment with open science infrastructures that support persistent identifiers, versioning, and long-term accessibility.

\begin{findingbox}{Finding 1}
The rapid growth of LLM4RE datasets since 2022 is accompanied by a shift from ad hoc, project-specific hosting toward standardized open science platforms, indicating increasing maturity in dataset dissemination and accessibility practices.
\end{findingbox}

\begin{figure}[h]
  \centering
  \includegraphics[width=\columnwidth]{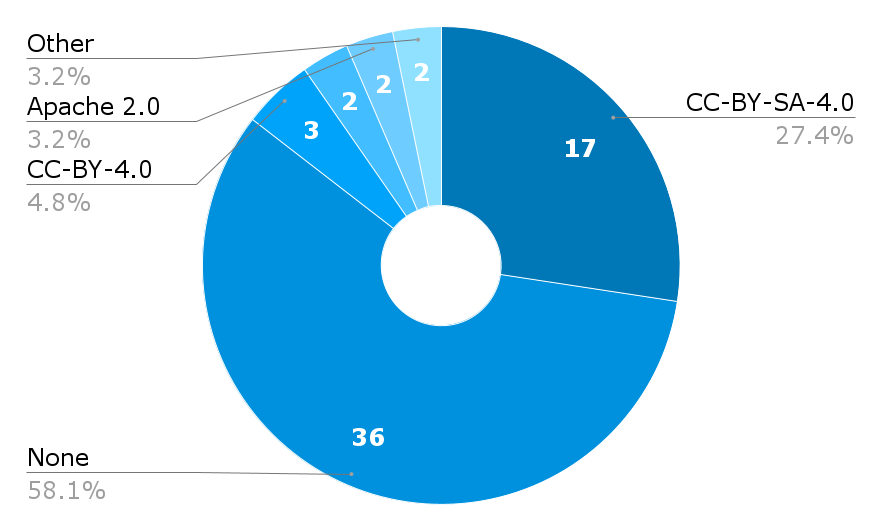}
  \caption{Distribution of dataset licensing.}
  \label{fig:licenses}
\end{figure}

With respect to licensing, Figure~\ref{fig:licenses} shows that more than half of the datasets (57.4\%) do not explicitly report a license. This issue persists in recent releases: among the 21 datasets published in 2025, 15 (71.4\%) lack an explicit public license.
When a license is specified, permissive and open licenses dominate, most notably Creative Commons licenses such as CC-BY-SA (27.9\%) and CC-BY-4.0 (4.9\%), alongside a smaller share of software-oriented licenses such as Apache~2.0 (3.2\%) and MIT (1.6\%) or even custom adaptations of MIT licensing such as SnT (1.6\%). 

\begin{findingbox}{Finding 2}
While the presence of open licenses supports reuse and evaluation, the high proportion of datasets without clear licensing information represents a significant obstacle for reproducibility, legal reuse, and systematic evaluation in LLM4RE research.
\end{findingbox}

Complementarily to this analysis, reuse across studies is limited: only 11 of the 62 datasets (17.7\%) are reused as-published in more than one study, with \texttt{PURE}~\cite{ferrari2018pure} and \texttt{PROMISE}~\cite{clelandhuang2007nfr} being the most reused datasets, appearing in five and four publications, respectively. 
On the other hand, 14 datasets (22.6\%) extend one or more previously released datasets. 
Similarly, \texttt{PROMISE}~\cite{clelandhuang2007nfr} and \texttt{PURE}~\cite{ferrari2018pure} are the most frequently extended datasets, with four and three direct extensions, respectively. In both cases, extensions also propagate beyond the original datasets through secondary derivations, such as \texttt{PROMISE\_exp}~\cite{lima2019promiseexp}, which augments \texttt{PROMISE} with additional requirements, and \texttt{PURE\_conflicts}~\cite{malik2023augmentation}, which extends \texttt{PURE} by introducing conflict labels between requirement pairs.
Beyond direct extensions, some of these datasets primarily act as aggregations of multiple prior datasets enriched with additional semantic information. Examples include \texttt{Cross-project\_trace}~\cite{ge2025crosslevel}, which combines several cross-project datasets annotated with requirement dependencies, and \texttt{Synthline\_conflicts}~\cite{synthline2024nli}, which aggregates heterogeneous sources to provide conflict annotations.

\begin{findingbox}{Finding 3}
Reuse in LLM4RE datasets is limited and highly concentrated: only a small subset is reused across multiple studies (17.7\%) or extended through derivative datasets (22.6\%), typically building on a few well-established resources (e.g., \texttt{PROMISE} and \texttt{PURE}), while the majority remain confined to single-study use.
\end{findingbox}
\section{Characterization (RQ2)}
\label{sec:rq2}

\begin{figure}[h]
  \centering
  \includegraphics[width=\columnwidth]{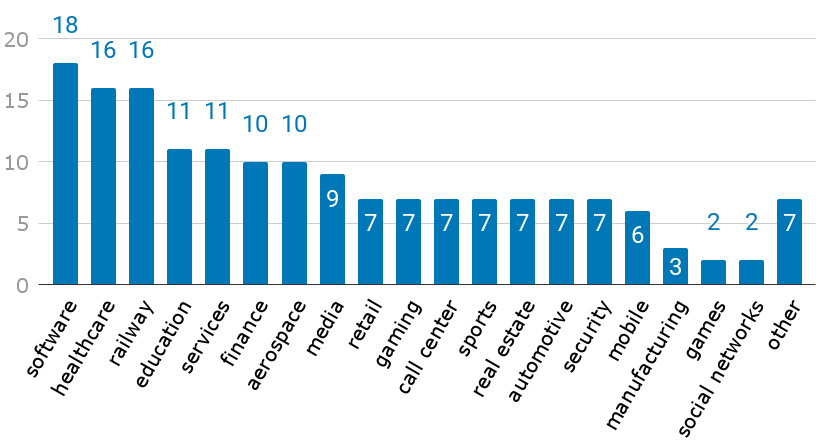}
  \caption{Distribution of domains.}
  \label{fig:domains}
\end{figure}

Figure~\ref{fig:domains} summarizes the application domains covered by the 62 datasets. Most datasets are single-domain (45), while a smaller but non-negligible subset spans multiple domains (17). Domain–dataset distribution is polarized, with multi-domain datasets clustering at two to three domains (7) and at twelve or more domains (7), and only a few datasets (3) spanning intermediate ranges. 
With respect to domain coverage, the corpus is led by software (18), with representative datasets such as \texttt{Acharya2024}~\cite{acharya2024bugzilla} and \texttt{SecReq}~\cite{wang2019secreq}, followed by healthcare (16), for example \texttt{CCHIT} and \texttt{InfusionPump}~\cite{ge2025crosslevel}, and railway (16), including \texttt{RailwayReq}~\cite{perko2024railway} and \texttt{ETCS\_conflicts}~\cite{fantechi2023inconsistency}. Less frequent domains rarely appear throughout the corpus, including legal requirements (e.g., \texttt{MURCIA}~\cite{abualhaija2024murcia}) and commerce (e.g., \texttt{LEXDEMOD}~\cite{sancheti2022lexdemod})

Regarding language coverage, the dataset landscape is overwhelmingly English-centric: 58 of the 62 datasets (93.5\%) are available exclusively in English, reflecting the dominant language of publication and evaluation in LLM4RE research. Only three datasets are written entirely in Chinese, namely \texttt{LERE}~\cite{han2025lere}, \texttt{MetroCode\_entities} and \texttt{MetroCode\_types}~\cite{metrocode2024}, which focus on error classification and information extraction from railway and industrial requirements. In addition, a single dataset explicitly supports multiple languages, \texttt{Wei2023}~\cite{wei2023bilingual}, which provides bilingual (English–French) app reviews to study zero-shot and cross-lingual LLM behavior. 

\begin{findingbox}{Finding 4}
The current LLM4RE dataset landscape supports evaluation across a wide range of application domains, including heterogeneous single-domain and explicitly multi-domain datasets, enabling partial assessment of domain generalization. However, this generalization capacity is confined to English-language settings, indicating that while domain transfer is increasingly feasible, language transfer remains a major open challenge.
\end{findingbox}

\begin{figure}[h]
  \centering
  \includegraphics[width=\columnwidth]{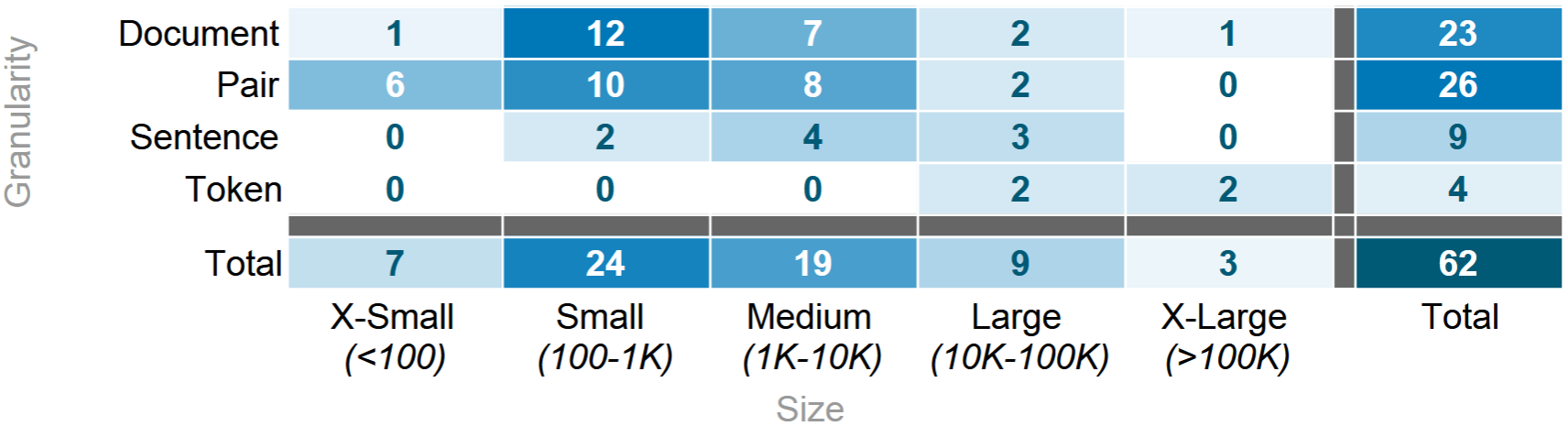}
  \caption{Distribution of dataset granularity and size.}
  \label{fig:granularity-size}
\end{figure}

Figure~\ref{fig:granularity-size} shows the distribution of datasets across artifact (i.e., document) granularity and size. Granularity denotes the unit of analysis, namely documents, sentences, tokens, or paired artifacts (e.g., requirement pairs with a dependency). Size corresponds to the number of such units in each dataset. For analytical clarity, dataset sizes are grouped into logarithmic categories following an order-of-magnitude ($\log_{10}$) progression to support comparative analysis; exact dataset cardinalities are reported in the public catalogue. 

Results highlight a clear concentration of LLM4RE datasets around high-granularity, mid-scale artifacts. Document-level corpora are predominantly small to medium (30.6\%), with comparatively few datasets above 10K instances. Typically, datasets remain limited in volume even when reused across studies (e.g., \texttt{PROMISE}~\cite{menzies2016promise} and its derivatives; and traceability datasets, e.g., \texttt{CCHIT}, \texttt{InfusionPump}~\cite{ge2025crosslevel}). Pair-based datasets exhibit a similar profile with small-to-medium collections (29.0\%), aligning with the higher cost of constructing labelled pairs for tasks such as conflict detection and traceability (e.g., \texttt{ETCS\_conflicts}~\cite{fantechi2023inconsistency}, \texttt{WorldVista\_conflicts})~\cite{malik2023augmentation}. In contrast, token-level resources appear only above 10K scales, as token annotation naturally inflates dataset cardinality, (e.g., \texttt{Motger2024}~\cite{motger2024tfrex}, \texttt{AutoFactory}~\cite{boudribila2025autofactory}). Finally, sentence-level datasets sit between extremes, clustering in medium to large sizes (14.5\%), consistent with sentence segmentation being a low-cost expansion of document collections and a common basis for classification benchmarks (e.g., \texttt{ReqEval}~\cite{yildirim2024reqeval}, \texttt{Unfair-TOS}~\cite{lippi2019claudette}). 

These results indicate a recurring trade-off between artifact granularity and dataset scale in LLM4RE datasets. Most resources target document- and pair-level units, which align well with common RE tasks (e.g., classification, traceability, and conflict detection) but require substantial manual effort, resulting predominantly in small to medium collections (e.g., \texttt{CCHIT} and \texttt{InfusionPump}~\cite{ge2025crosslevel}, \texttt{ETCS\_conflicts}~\cite{fantechi2023inconsistency}). In contrast, the few large to very large datasets arise either because artifacts are naturally abundant (e.g., \texttt{Acharya2024} bug reports~\cite{acharya2024bugzilla}) or because token-level representations inflate cardinality once annotation is defined (e.g., \texttt{Motger2024} \cite{motger2024tfrex} and \texttt{AutoFactory}~\cite{boudribila2025autofactory}). Consequently, benchmark coverage is broad in task and domain labels, but uneven with respect to dataset volume and granularity.

\begin{findingbox}{Finding 5}
LLM4RE evaluation is dominated by small to medium document- and pair-level datasets, while evidence at large scale is concentrated in a few high-volume or token-level corpora. As a result, model performance is well studied across tasks and domains but remains sparsely assessed under combined high-scale and fine-grained conditions.
\end{findingbox}
\section{Applicability in RE (RQ3)}
\label{sec:rq3}

\begin{figure}[h]
  \centering
  \includegraphics[width=\columnwidth]{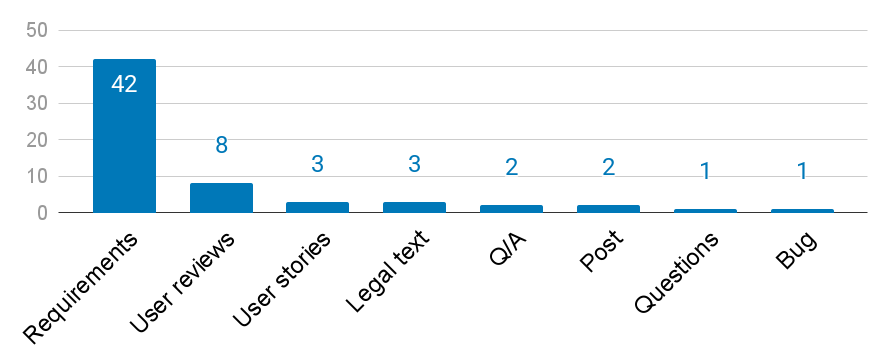}
\caption{Distribution of artifacts.}
\label{fig:artifacts}
\end{figure}

Figure~\ref{fig:artifacts} shows that variability in artefact type across datasets is low. Most datasets focus on requirements expressed as structured or semi-structured documents (67.7\%), reflecting the central role of requirements specifications in LLM-based RE research (e.g., \texttt{PROMISE}~\cite{menzies2016promise}). Variant representations (e.g., user stories) are only marginally present (4.8\%) (e.g., \texttt{Bragilovski2024}~\cite{bragilovski2024userstories}). User feedback is the second most common artefact type, mainly through user reviews (12.9\%) and, to a lesser extent, online forum posts (3.2\%), indicating a growing but still limited use of post-deployment feedback in RE tasks (e.g., \texttt{AWARE}~\cite{alturaief2021aware}). Other artefact types appear only sporadically, including legal or regulatory texts (4.8\%) (e.g., \texttt{Azeem2023}~\cite{azeem2024gdpr}), question--answer (Q/A) pairs (3.2\%) (e.g., \texttt{AeroReqEval}~\cite{yang2025aeroreqeval}), and competency questions (1.6\%) (e.g., \texttt{CQ-BEN}~\cite{alharbi2024cqbench}). 

\begin{figure}[t]
  \centering
  \includegraphics[width=\columnwidth]{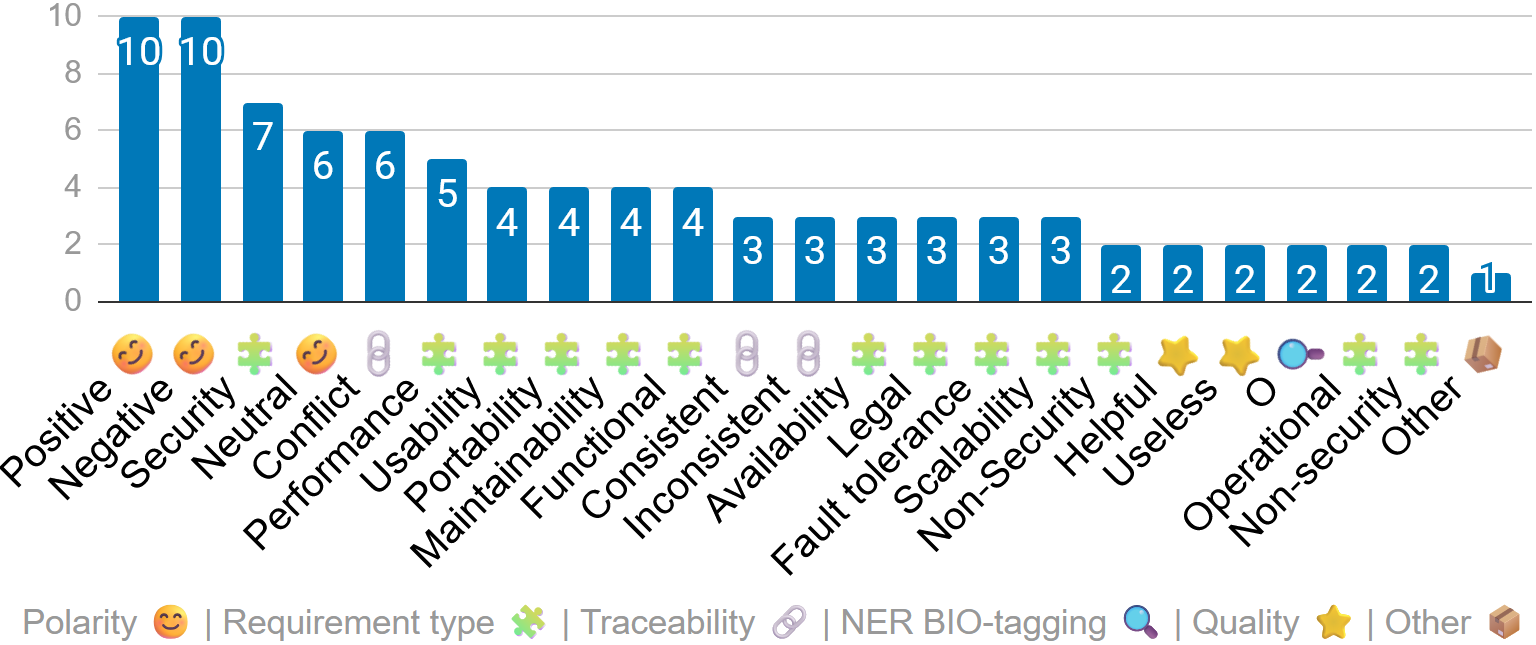}
\caption{Distribution of labels across datasets.}
\label{fig:labels}
\end{figure}

Figure~\ref{fig:labels} reports the distribution of annotation labels across datasets. Overall, 46 datasets (74.2\%) provide explicit labels, while the remainder are released as unlabelled corpora used mainly for retrieval, prompting, or qualitative evaluation. The most frequent label family concerns polarity in user feedback, with \emph{Positive} and \emph{Negative} (and less frequently \emph{Neutral}) dominating review-centric datasets (e.g., \texttt{AWARE}~\cite{alturaief2021aware}). A second prominent pattern captures requirements types, especially non-functional requirement categories, such as \emph{Security}, \emph{Performance}, and \emph{Usability}, which are widely used in requirements classification benchmarks (e.g., \texttt{PROMISE} and \texttt{PROMISE\_exp}~\cite{menzies2016promise}). In management-oriented datasets, labels often encode pairwise relations for traceability and consistency, where conflict detection is typically framed as \emph{Conflict}/\emph{Neutral} or \emph{Consistent}/\emph{Inconsistent} (e.g., \texttt{PURE\_conflicts} and \texttt{ETCS\_conflicts}~\cite{malik2023augmentation,fantechi2023inconsistency}). 
Finally, less represented annotations include token-level sequence labels using BIO-style tags for requirement components or entities, with the \texttt{O} tag for non-target tokens (e.g., \texttt{AutoFactory}~\cite{boudribila2025autofactory}, \texttt{PET}~\cite{bellan2022pet}), as well as coarse-grained quality assessments of document usefulness such as \emph{Helpful}/\emph{Useless} (e.g., \texttt{Groen2025\_P1}~\cite{groen2025quality}).

\begin{findingbox}{Finding 6}
LLM4RE datasets are strongly centred on traditional requirements artefacts and their associated annotation schemes. Most resources focus on structured or semi-structured requirements documents and adopt labels tailored to established RE tasks, such as requirements type classification, sentiment polarity, and pairwise traceability or conflict detection, while alternative artefacts and annotation perspectives remain comparatively under-represented.
\end{findingbox}


\begin{figure}[b]
  \centering
  \includegraphics[width=\columnwidth]{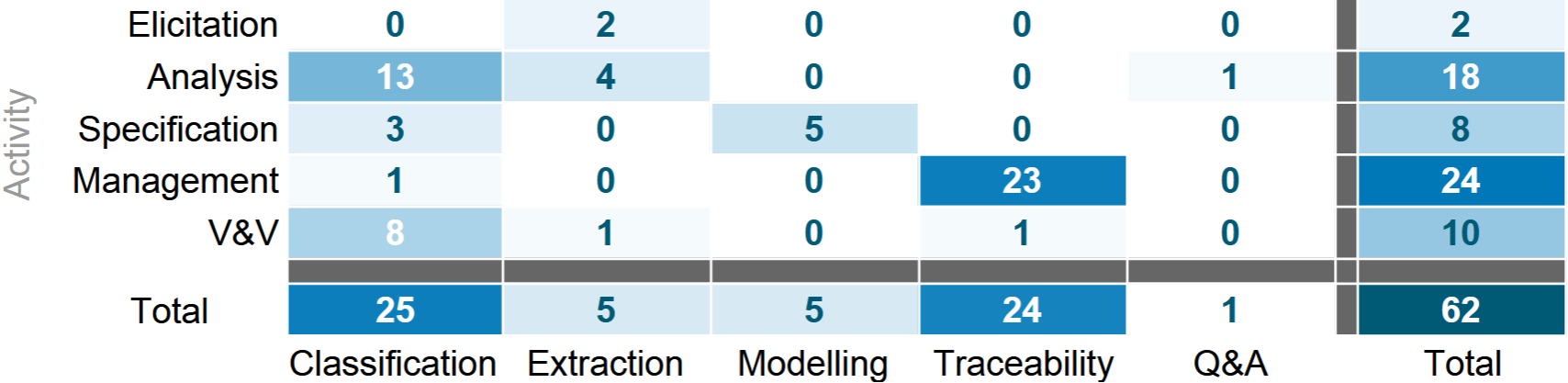}
\caption{Distribution across RE activities and NLP tasks.}
\label{fig:activity-task}
\end{figure}

Figure~\ref{fig:activity-task} illustrates the distribution of datasets across RE activities and tasks, using the vocabulary as defined in Section~\ref{sec:method}.


\textbf{Elicitation}. Only two datasets within the catalogue are used to support or assess requirements elicitation tasks: \texttt{CQ-BEN}~\cite{alharbi2024cqbench} and Dronology~\cite{clelandhuang2018dronology}. \texttt{CQ-BEN} operationalises elicitation by representing requirements as competency questions, capturing stakeholders' information needs and intended system scope~\cite{alharbi2024cqbench}. \texttt{Dronology} supports elicitation through a comprehensive set of stakeholder-driven artefacts, including goals, requirements, and environmental assumptions, enabling the study of requirements discovery in a realistic, safety-critical cyber-physical system context~\cite{clelandhuang2018dronology}. However, these datasets primarily capture post-elicitation artefacts, rather than empirical data from early elicitation activities such as interview transcripts, workshop records, or negotiation logs.
    
\textbf{Analysis}. A substantial portion of the catalogue (29.0\%) supports requirements analysis by enabling tasks such as classification, extraction, interpretation, and question answering over existing requirement artefacts. Representative datasets include \texttt{PROMISE}~\cite{menzies2016promise} and its extensions, which operationalise analysis through functional and non-functional classification, as well as domain-specific corpora such as \texttt{SecReq}~\cite{wang2019secreq}, \texttt{LEXDEMOD}~\cite{sancheti2022lexdemod}, and \texttt{MetroCode\_types}~\cite{metrocode2024}, which focus on semantic categorisation of security, legal, or regulatory requirements. Other datasets, such as \texttt{AeroReqEval}~\cite{yang2025aeroreqeval} and \texttt{MultiDomainFPA}~\cite{zhao2024fpa}, support analytical reasoning via question–answer pairs, while other like \texttt{PET}~\cite{bellan2022pet} and \texttt{Dalpiaz2018}~\cite{dalpiaz2018userstories} enable structural extraction from requirements expressed as process descriptions or user stories. Hence, analysis tasks are centred on refining and structuring requirements after their initial formulation.
    
\textbf{Specification}. Datasets in this category emphasise the formalisation and structuring of requirements into precise, model-oriented representations. Resources such as \texttt{AutoFactory}~\cite{boudribila2025autofactory} annotate requirement texts with fine-grained semantic components to support information extraction and automation. Other datasets, including \texttt{Guidotti2024}~\cite{guidotti2024psp}, \texttt{RailwayReq}~\cite{perko2024railway}, \texttt{Yilongfei2024\_1} and \texttt{Yilongfei2024\_2}~\cite{xu2024nltoltl}, operationalise specification through the translation of natural language requirements into structured or formal models. Complementarily, \texttt{Bragilovski2024}~\cite{bragilovski2024userstories} and \texttt{LERE}~\cite{han2025lere} support specification by enabling modelling from user stories or by identifying specification-level defects. These datasets reflect specification activities focused on increasing precision, structure, and formal analysability of requirements.
    
\textbf{Management}. As the most represented activity (38.7\%), datasets mapped to management predominantly support requirements traceability, consistency management, and change-related reasoning through paired requirement representations. A large subset of the catalogue consists of cross-level traceability datasets, such as \texttt{CCHIT}, \texttt{CM1}, \texttt{Modis}, \texttt{InfusionPump}, and \texttt{WARC}~\cite{ge2025crosslevel}, which link high-level and low-level requirements across domains. Complementarily, several datasets focus on conflict detection, including project-specific and cross-domain collections (e.g., \texttt{ETCS/Library\_conflicts}~\cite{fantechi2023inconsistency}), as well as larger-scale datasets derived from industrial or community sources (e.g., \texttt{Bugzilla\_conflicts}~\cite{malik2023augmentation}).

\textbf{Verification and validation}. Datasets mapped to verification and validation primarily support the assessment of requirement quality, completeness, and alignment with stakeholder feedback through classification, extraction, and traceability tasks. Several datasets leverage user feedback to validate requirements against real-world usage and expectations, including \texttt{AWARE}~\cite{alturaief2021aware}, the collections from \texttt{Groen2025} collections~\cite{groen2025quality}, \texttt{Khan2025}~\cite{khan2025emotions}, and \texttt{PCTD}~\cite{zhang2023mining}, which enable the evaluation of relevance, quality aspects, or emotional signals in app reviews. Other datasets focus on validating requirements through software-related artefacts, such as bug reports (\texttt{Acharya2024}~\cite{acharya2024bugzilla}) or legal and regulatory texts (\texttt{Azeem2023}~\cite{azeem2024gdpr}). 

\begin{findingbox}{Finding 7}
LLM4RE datasets are unevenly distributed across the RE lifecycle and NLP tasks: management and analysis dominate the landscape, while elicitation is scarcely represented. This suggests that current dataset support is strongest for post-elicitation activities and weakest for early requirements discovery. Complementarily, dataset availability is skewed toward tasks well suited to post-hoc annotation of existing artefacts (e.g., classification and traceability), while tasks that better support exploratory and early-stage RE (e.g., Q\&A), remain underrepresented.
\end{findingbox}

\section{Research Gaps and Opportunities (RQ4)}
\label{sec:rq4}

Based on the empirical findings from RQ1–RQ3, we identify several key gaps in the current LLM4RE dataset landscape and propose corresponding opportunities for future work:

\begin{gapoppbox}
\textcolor{red!75!black}{\textbf{\ding{55} Gap:}} Dataset governance practices do not reflect the role of LLM4RE datasets as long-lived research infrastructure (e.g., unclear licensing, transient hosting).

\textcolor{green!60!black}{\textbf{\ding{52} Opportunity:}} Standardize dataset documentation and archiving (explicit licenses, persistent repositories) to improve longevity and legal reusability.
\end{gapoppbox}

Only a subset of LLM4RE datasets adhere to open-science standards. Over half the identified datasets lack any explicit license, and several surveyed studies had to be excluded due to broken links or inaccessible data. While hosting practices are gradually consolidating on stable platforms such as GitHub and Zenodo, many datasets remain on short-lived infrastructures. This limits reproducibility, inhibits legal reuse, and prevents datasets from serving as reliable benchmarks over time. Enforcing common metadata schemas (e.g., SPDX licensing) and publishing datasets on archival platforms with DOIs would help position datasets as shared, durable infrastructure for LLM4RE research.

\begin{gapoppbox}
\textcolor{red!75!black}{\textbf{\ding{55} Gap:}} Dataset artefacts predominantly reflect post-hoc requirements, not the stakeholder interaction and rationale that produces them.

\textcolor{green!60!black}{\textbf{\ding{52} Opportunity:}} Release stakeholder-grounded datasets that preserve intent, negotiation, and justification signals in privacy-aware forms.
\end{gapoppbox}

Although RQ3 identifies a small set of elicitation-related datasets, these mainly capture post-elicitation artefacts rather than empirical traces of elicitation itself, such as interviews, workshops, or trade-off discussions. As a result, LLMs are primarily evaluated as annotators of already-written artefacts, rather than as assistants that support requirements discovery, clarification, and justification. A promising direction is to curate datasets that include stakeholder utterances or questions, intermediate representations (e.g., candidate requirements, assumptions, open issues), and decision rationales explaining why formulations were accepted, rejected, or deferred. Even lightweight annotations, such as speaker role, intent type, or decision outcome, would enable richer evaluation of LLM capabilities that are central to RE.

\begin{gapoppbox}
\textcolor{red!75!black}{\textbf{\ding{55} Gap:}} Current benchmarks rarely capture end-to-end RE workflows across multiple lifecycle activities.

\textcolor{green!60!black}{\textbf{\ding{52} Opportunity:}} Create scenario-based datasets that link heterogeneous artefacts across activities to evaluate LLMs performance holistically.
\end{gapoppbox}

RQ3 highlight the absence of datasets that preserve continuity across the RE lifecycle. This fragmentation limits evaluation in realistic settings, where practitioners iterate across multiple lifecycle activities. A concrete direction is to design ``threaded'' benchmarks in which a single scenario links heterogeneous artefacts, such as stakeholder questions, evolving requirements, structured models, trace links, and validation from user feedback. These datasets can be constructed by aligning existing artefact sources already present in the catalogue, enabling cross-activity evaluation rather than isolated single-task benchmarks.

\begin{gapoppbox}
\textcolor{red!75!black}{\textbf{\ding{55} Gap:}} Domain diversity is growing, but language and socio-technical interaction diversity remain largely unrepresented.

\textcolor{green!60!black}{\textbf{\ding{52} Opportunity:}} Establish multilingual, locale-aware benchmarks with aligned artefacts and evaluation splits for cross-lingual and cross-context transfer.
\end{gapoppbox}

Figure~\ref{fig:domains} shows that datasets span multiple domains, yet language coverage remains overwhelmingly English, as highlighted in RQ2, with only a few non-English or bilingual datasets. This limits the external validity of generalization claims, since models are rarely evaluated under realistic multilingual conditions where requirements involve local terminology, regulatory phrasing, and culturally shaped ambiguity. The opportunity goes beyond translating existing corpora and lies in building multilingual datasets that preserve RE phenomena, such as parallel artefacts for the same scenario, locale-specific constraints, and evaluation splits that allow domain transfer effects to be examined independently from language transfer. While initial examples exist in the catalogue, a cohesive benchmark suite could standardise cross-lingual evaluation and better expose language- and domain-specific model limitations.

\begin{gapoppbox}
\textcolor{red!75!black}{\textbf{\ding{55} Gap:}} Dataset scale and packaging practices limit reproducibility and restrict experimentation beyond small, single-task evaluations.

\textcolor{green!60!black}{\textbf{\ding{52} Opportunity:}} Release larger, structured, and reusable corpora with standard metadata, splits, and evaluation protocols for modern LLM workflows (e.g., RAG, tool-augmented agents).
\end{gapoppbox}

Figure~\ref{fig:granularity-size} indicates that most datasets are small to medium and concentrated at document- and pair-level granularity, which aligns with common RE tasks but constrains training, benchmarking, and retrieval-based experimentation. At the same time, RQ1 shows that dissemination is maturing (Figure~\ref{fig:source}), yet licensing remains a major barrier (Figure~\ref{fig:licenses}). A practical, high-leverage opportunity is to treat datasets as reusable infrastructure: publish machine-readable metadata (including SPDX-compatible licensing), stable identifiers, versioned releases, and standard splits that control train-test data leakage. Structurally, datasets should expose hierarchy and context (project, component, linked artefacts) to support RAG and agentic workflows, rather than flattening artefacts into isolated instances.
\section{Discussion}
\label{sec:discussion}

\subsection{Contributions and Implications}
\label{sec:contribution-implications}

This study consolidates evidence from 62 datasets to characterise the current LLM4RE dataset landscape and to derive actionable implications for dataset creators, tool builders, and evaluators. By combining a structured characterisation schema (RQ2), provenance and accessibility analysis (RQ1), and RE lifecycle applicability mapping (RQ3), we provide both a descriptive snapshot of what exists and high-level suggestions for how future datasets can be documented, reused, and evaluated more systematically.

\textbf{\(\triangleright\) Implications for dataset documentation and reuse.} A core contribution of this mapping study is the dataset characterisation scheme (Table~\ref{tab:extraction}), designed to be reusable beyond this paper for both (i) systematic mapping updates and (ii) dataset documentation by authors. Unlike paper-centric reporting, Table~\ref{tab:extraction} defines dataset-centric descriptors directly relevant to reuse, including granularity, size, language, task framing, hosting, and licensing. This reinforces a key takeaway from RQ1: while dissemination practices are becoming more stable (Figure~\ref{fig:source}, Finding~1), missing or unclear licensing remains a concrete barrier to legal reuse and reproducibility (Figure~\ref{fig:licenses}, Finding~2). Consequently, even widely used datasets may become effectively unusable over time.

To mitigate this, we advocate treating RE datasets as long-lived research infrastructure rather than supplementary artefacts. At minimum, datasets should be released with (i) explicit licenses, preferably declared in a machine-readable form (e.g., SPDX identifiers), (ii) persistent archiving (e.g., Zenodo DOIs), and (iii) versioned releases aligned with the artefact evolution that is common in RE. Such practices directly support reproducibility and also enable cumulative science, where later work can build on a stable base rather than repeatedly re-curating similar resources.

\textbf{\(\triangleright\) Catalogue and knowledge graph as complementary research artefacts.} A second contribution is the combination of two complementary representations of the dataset landscape: a human-oriented catalogue and a machine-actionable ORKG comparison. Together, these forms balance exploratory analysis (e.g., manual inspection, filtering, and qualitative comparison) with computational reuse (e.g., querying by task, granularity, licensing, or language) and future extensibility. In particular, a knowledge graph representation supports automated meta-analyses and can facilitate continuous updates as new datasets appear, which is crucial given the post-2022 growth trend observed in Figure~\ref{fig:year-distribution}. This aims to lower the effort of repeating similar mapping studies and to support dataset curation as a shared, evolving community resource.

\textbf{\(\triangleright\) Reframing ``dataset applicability'' beyond current usage.} RQ3 maps datasets to RE activities and tasks based on how they are used in the literature, revealing strong imbalances and tight coupling between activities and NLP task framings (Figure~\ref{fig:activity-task}, Finding~7). While these results accurately reflect current practice, they should not be interpreted as hard limits on what the datasets can support. Many datasets can be repurposed, either as-is or via lightweight relabelling, re-segmentation, or pairing strategies. For example, document-level corpora used for classification can often be re-expressed as (i) sentence-level datasets for finer-grained evaluation, (ii) retrieval corpora for RAG setups, or (iii) pair-based datasets via constructed links (e.g., requirement-to-rationale, requirement-to-test, or requirement-to-feedback).

This distinction matters because it changes how we read the ``dominance'' of management and analysis. The current landscape is skewed toward tasks that are relatively straightforward to label and benchmark (e.g., traceability pairs and classification), but the underlying artefacts frequently contain richer structure that could support broader RE workflows. A practical implication is that future dataset papers should report not only the primary task, but also plausible secondary uses (and recommended splits) to encourage cross-task reuse and reduce redundant dataset creation.

\textbf{\(\triangleright\) From benchmark collections to research infrastructure.} Taken together, our results suggest that the main limitation of the current landscape is not a lack of datasets, but a lack of composability. Datasets are typically released as single-purpose benchmarks, tightly coupled to one task framing and usage context, which makes it difficult to assemble realistic, end-to-end evaluation scenarios. This matters because LLM4RE practice increasingly relies on multi-step pipelines, such as retrieval, decomposition, and iterative refinement, where performance depends on integrating heterogeneous evidence and maintaining consistency across artefacts rather than on isolated accuracy. The gaps identified in RQ4 therefore point to a shift in how resources should be designed and published, toward datasets that are (i) linkable across artefact types and lifecycle activities, (ii) portable across contexts, including multilingual and locale-aware settings, and (iii) maintainable over time through versioned releases with traceable evolution, in addition to being openly reusable via clear licensing and persistent archiving. In this sense, the catalogue and ORKG comparison are not only reporting devices but enabling mechanisms, as they provide the structure needed to discover, connect, and repurpose datasets beyond their original task definitions, supporting cumulative benchmark construction instead of repeated, isolated dataset creation.

\subsection{Threats to validity}
\label{sec:ttv}
We structure threats to validity following Wohlin et al.~\cite{Wohlin2012}, considering construct, internal, external, and conclusion validity.

\textbf{Construct validity.}
Construct validity arises from (i) normalising heterogeneous dataset descriptions into a unified extraction schema (Table~\ref{tab:extraction}), (ii) assigning domains, artefact granularities, and size bins, and (iii) mapping datasets to RE activities and NLP task types (RQ3). Although our coding builds on established RE lifecycle terminology and prior taxonomies (Section~\ref{sec:method}), some datasets are multi-purpose or are used in ways that diverge from their original intent, which can yield borderline cases and alternative plausible mappings. Likewise, discretising continuous properties (e.g., dataset size) into bands may hide nuances near thresholds, even when exact cardinalities are preserved. We mitigate these by retaining original metadata in the replication package, and by using normalised labels extensively used in related work only for aggregation and comparative analyses, enabling re-coding under alternative schemes.

\textbf{Internal validity.}
Internal validity threats emerge from selection and extraction bias during screening and coding, including errors caused by incomplete reporting in primary studies. We mitigated these risks through a predefined protocol, independent dual screening, and reconciliation, achieving high inter-rater agreement. Remaining risks include (i) misclassification when dataset documentation omits key properties, and (ii) survivorship bias effects where inaccessible datasets are less likely to be inspected in depth. We partially address this by explicitly reporting accessibility and licensing as first-class outcomes (RQ1) and by making extraction decisions transparent in the replication package.

\textbf{External validity.}
Our findings are constrained by the selected search strategy, including the definition of the search string and selected sources. Relevant datasets may exist outside the selected databases, including grey literature, community repositories, and industry disclosures, which is increasingly plausible given the rapid pace of LLM research and the role of repository-first dissemination (e.g., Hugging Face). Consequently, our catalogue should be interpreted as a structured snapshot of the indexed research landscape rather than an exhaustive census of all available LLM4RE datasets. As continuation work, we plan to extend discovery through broader queries, backward and forward snowballing, and systematic inclusion of grey literature and repository-centric search.

\textbf{Conclusion validity.}
Conclusion validity stems from the nature of the study and the heterogeneity of dataset reporting: quantitative trends (e.g., distributions across activities, tasks, or size bands) may be sensitive to a small number of influential datasets or to alternative coding choices for borderline cases. In addition, accessibility is time-dependent: links may break, repositories may move, and licenses may be updated after our data collection, which can affect conclusions about availability and reuse constraints. We mitigate these threats by (i) reporting results as descriptive rather than causal claims, and (ii) releasing the catalogue and ORKG comparison to support independent verification and re-analysis. Nevertheless, statements about accessibility and licensing should be interpreted as a snapshot at the time of data collection.

\section{Conclusions and Future Work}
\label{sec:conclusions}

This paper presented a systematic mapping study of publicly available datasets used in LLM4RE research. By analysing 45 primary studies and characterising 62 datasets through a structured extraction schema, we provide an evidence-based view of how LLM4RE research is shaped by dataset availability, packaging, and coverage. We observe rapid growth in dataset usage after 2022 and increased adoption of open-science hosting platforms, yet reuse remains limited by unclear licensing and time-dependent accessibility. The landscape is also uneven across the RE lifecycle: datasets mainly support management and analysis, artefact types and annotations are centred on traditional requirements documents and task-specific label schemes, and language and socio-technical diversity remain limited. These patterns help explain why many LLM4RE evaluations remain task-local and difficult to compare across lifecycle activities or deployment contexts. Beyond these findings, we contribute reusable community infrastructure, including a dataset characterisation scheme and openly available artefacts that make the dataset landscape searchable, comparable, and extensible (replication package, web catalogue, and ORKG comparison).

As future work, we plan to (i) extend surveyed studies via snowballing and inclusion of grey literature and repository-centric search, (ii) maintain the catalogue as a living resource with versioned snapshots, and (iii) enrich metadata to better support dataset composability (e.g., lineage, recommended splits, and reuse constraints). Finally, building on the gaps identified in RQ4, we aim to curate scenario-based benchmark configurations that connect heterogeneous artefacts across lifecycle activities, enabling evaluation of modern LLM4RE workflows (e.g., RAG and multi-step settings) rather than isolated single-task benchmarks.

\section*{Data Availability Statement}
\label{sec:data}

A replication package including the search study results, screening procedure, and data extraction 
is permanently hosted on Zenodo: \href{https://doi.org/10.5281/zenodo.18258344}{https://doi.org/10.5281/zenodo.18258344}.


\clearpage
\bibliographystyle{ACM-Reference-Format}
\bibliography{ref}

\end{document}